\documentclass[aps,prb,twocolumn,floatfix,superscriptaddress,showpacs,psfig,reprint, longbilbiography]{revtex4-2}
\usepackage{amsmath}
\usepackage{amssymb}
\usepackage{amsthm}
\usepackage{amsfonts}
\usepackage[version=4]{mhchem}
\usepackage{listings}
\usepackage{enumerate}
\usepackage{latexsym}
\usepackage{psfrag}
\usepackage{bm} 
\usepackage[all]{xy}
\usepackage{graphicx}
\usepackage{subfigure}
\usepackage[pdftex,colorlinks]{hyperref}
\usepackage{color}
\usepackage{mathtools}
\usepackage{times}
\usepackage{array}
\usepackage{placeins}
\usepackage{comment}

\hypersetup{
    colorlinks=true,
    linkcolor=blue,
    filecolor=blue,
    urlcolor=blue,
    citecolor=blue,
}

\begin{document}
\title{Robust superconducting correlation against inter-site interactions
in the extended two-leg Hubbard ladder }
\author{Zongsheng Zhou}
\affiliation{School of Physical Science and Technology \& Key Laboratory for Magnetism
and Magnetic Materials of the MoE, Lanzhou University, Lanzhou 730000, China.}
\affiliation{Beijing National Laboratory for Condensed Matter Physics and Institute of 
Physics, Chinese Academy of Sciences, Beijing 100190, China}
\affiliation{Lanzhou Center for Theoretical Physics and Key Laboratory of Theoretical
Physics of Gansu Province, Lanzhou University, Lanzhou 730000, China.}
\author{Weinan Ye}
\affiliation{School of Physical Science and Technology \& Key Laboratory for Magnetism
and Magnetic Materials of the MoE, Lanzhou University, Lanzhou 730000, China.}
\affiliation{Lanzhou Center for Theoretical Physics and Key Laboratory of Theoretical
Physics of Gansu Province, Lanzhou University, Lanzhou 730000, China.}
\author{Hong-Gang Luo}
\affiliation{School of Physical Science and Technology \& Key Laboratory for Magnetism
and Magnetic Materials of the MoE, Lanzhou University, Lanzhou 730000, China.}
\affiliation{Lanzhou Center for Theoretical Physics and Key Laboratory of Theoretical
Physics of Gansu Province, Lanzhou University, Lanzhou 730000, China.}
\affiliation{Beijing Computational Science Research Center, Beijing 100084, China}
\author{Jize Zhao}
\email{zhaojz@lzu.edu.cn}
\affiliation{School of Physical Science and Technology \& Key Laboratory for Magnetism
and Magnetic Materials of the MoE, Lanzhou University, Lanzhou 730000, China.}
\affiliation{Lanzhou Center for Theoretical Physics and Key Laboratory of Theoretical
Physics of Gansu Province, Lanzhou University, Lanzhou 730000, China.}
\author{Jun Chang}
\email{junchang@snnu.edu.cn}
\affiliation{College of Physics and Information Technology, Shanxi Normal University, 
Xi’an 710119, China}
 
\begin{abstract}
The Hubbard and related models serve as a fundamental starting point
in understanding the novel experimental phenomena in correlated electron
materials, such as superconductivity, Mott insulator, magnetism and
stripe phases. Recent numerical simulations indicate that the emergence
of superconductivity is connected with the next nearest-neighbor hopping
$t^{\prime}$ in the Hubbard model. However, the impacts of complex
inter-site electron interaction in the $t^{\prime}$-Hubbard model
are less explored. Utilizing the state-of-art density-matrix renormalization
group method, we investigate the $t^{\prime}$-Hubbard model on a
two-leg ladder with inter-site interactions extended to the fourth neighbor
sites. The accurate numerical results show that the quasi-long-range
superconducting correlation remains stable under the repulsive nearest-neighbor
and the next nearest-neighbor interactions though these interactions
are against the superconductivity. The ground state properties are
also undisturbed by the longer-range repulsive interactions. In addition,
inspired by recent experiments on one-dimensional cuprates chain ${\rm {Ba}_{2-x}{\rm {Sr}_{x}{\rm {CuO}_{3+\delta}}}}$,
which implies an effective attraction between the nearest neighbors
may exist in the cuprates superconductors, we also show that the attractive
interaction between the nearest neighbors significantly enhances
the superconducting correlation when it is comparable to the strength
of the nearest-neighbor hopping $t$. Stronger attraction drives the system into a 
Luther-Emery liquid phase. Nevertheless, with the attraction further increasing, the system enters an electron-hole 
phase separation and the superconducting correlation is destroyed. Finally, we investigate the effects
of on-site Coulomb interaction on superconductivity. 
\end{abstract}
\maketitle
\section{Introduction}

Despite the extensive investigations in the last decades, the microscopic
mechanism of high-$T_{c}$ superconductivity in cuprates remains one
of the puzzles in condensed matter physics~\cite{Keimer2015,Proust2019}.
Single band Hubbard model~\cite{Hubbard1963} and $t$-$J$ model~\cite{ZhangFC1988}
are the simplest models frequently employed to understand experimental
results in the high-$T_{c}$ cuprates. The latter is the strong interaction
limit of the former under hole doping, and is also the low energy 
effective model of the original three band $d$-$p$ model~\cite{EmeryVJ1987}, which
directly depicts the physics on the $\mathrm{CuO}_{2}$ plane in high-$T_{c}$
cuprates. In the single band Hubbard model on a square lattice, many
phases observed in high-$T_{c}$ cuprates are reproduced, such as
the antiferromagnetic magnetism at half filling~\cite{Hirsh1989,WhiteSR1989,QinMinpu2016}
as well as the antifferomagnetic correlation upon hole doping~\cite{WietekA2021}, pseudogap~\cite{GullE2013,WuWei2018,WietekA2021,HilleC2020}, 
the stripe phase
where charge density wave (CDW) and spin density wave (SDW) coexist
around optimal doping~\cite{EhlersG2017,QinMP2020} and metal phase under overdoping~\cite{KungYF2015}.
It means that the Hubbard model captures some of the crucial ingredients
of high-$T_{c}$ cuprates.

Recently, it was found that the Hubbard model with an attractive
 nearest-neighbor (NN) interaction could well explain the results of angle-resolved 
photoemission  spectroscopy on a one dimensional cuprates chain compound ${\rm {Ba}}_{2-x}{\rm {Sr}}_{x}{\rm {CuO}}_{3+\delta}$
~\cite{ZhuoyuCHen2021}. Due to various cuprates chain and ladder materials, the Hubbard related models on chain and ladders have been extensively
studied~\cite{Kuroki1996,Balents1996,Noack1996,DaulS2000,LouisKim2001,Tsuchiizu2002,Degli2016,Nocera2018,LucaF2019,JiangHongchen2020,Gannot2020},
which could help us understand the related experiments and gain insights into
strongly correlated electron systems. However, for the simplest
Hubbard model with a single orbital, the NN hopping along with on-site Coulomb interaction, current
powerful numerical methods demonstrated that the ground
state is not the superconducting state~\cite{BoxiaoZheng2017,QinMP2020},
but a stripe phase with the wavelength of charge density $\lambda_{c}=8$,
which is recently observed in experiment~\cite{SDEdkins2019}. Meanwhile,
numerical results indicate the superconducting state is a high-energy
excitation state and several stripe phases are highly competitive
near the ground state~\cite{Corboz2011,EhlersG2017,BoxiaoZheng2017,Darmawan2018,IdoKota2018,Vanhala2018}.
These results imply that some crucial ingredients leading to the superconductivity
are missing in the simplest Hubbard model. Some neglected terms, such
as the hopping term beyond the NN sites and the inter-site
interactions among electrons, should be taken into consideration to
describe the physics of high-$T_{c}$ cuprates. Indeed, the next-nearest
neighbor(NNN) hopping term $t^{\prime}$ brings impressive change into the 
ground state, it does not only choose the wavelength of $\lambda_{c}=4$
stripe state as the ground state~\cite{Ponsioen2019} that is more
widely observed in experiments~\cite{Tranquada1995,Tranquada2004,YKohsaka2007},
but also induce quasi-long ranged superconducting correlation on
a four-leg cylinder~\cite{HongchenJiang2019,ChungChiaMin2020,JiangYifan2020}.
Very recently, numerical simulations indicate that the $t^\prime$-Hubbard model
is adaptable in qualitatively capturing the physics in high-$T_c$ cuprates~\cite{HaoXu2023}.

In actual materials, it is worthy of noting that
in one and two dimensions, the Coulomb screening is relatively weaker than
that in three dimensions. Therefore, it is probably hard to thoroughly
screen the long-range interactions among electrons into on-site interaction,
thus inter-site interactions are still considerable. And the hopping 
$t^{\prime}$ between the NNN sites can not be neglected  directly either.
Therefore, to have a better insight into the physics in cuprates, 
it is natural to consider the inter-site interactions in the 
$t^{\prime}$-Hubbard model~\cite{HirayamaM2018,HirayamaM2019}. The 
extended $t^\prime$-Hubbard on ladders should be a more reasonable 
entrance in understanding the cuprates ladder materials exhibiting 
superconductivity~\cite{Uehara1996,JHSchon2001,Fujiwara2005}. 
Firstly, ladders serve as a bridge between one-dimensional 
and two-dimensional systems, in some cases, ladder share some similar physics
with its two-dimensional counterpart. Secondly, due to the chemical similarities,
a deep understanding on the cuprates ladder systems can also aid in comprehending
the physics in two dimensional cuprates materials.
Generally, we consider repulsive inter-site interactions in the extended 
Hubbard model as they originate from the Coulomb interaction. The 
situation may change when there exists a process for electrons exchanging
virtual bosons, e.g. the electron-phonon coupling. For instance, the numerical
simulation on the Holstein-Hubbard chain indicates that an effective
attractive NN interaction can be mediated by long-range electron-phonon
interaction~\cite{WangYao2021}. Thus, it is also worthwhile exploring
the $t^{\prime}$-Hubbard model with attractive NN interaction.

Unearthing new contained physics and hunting for superconductivity
in the Hubbard model are important research topics. Though the extended Hubbard
model has been investigated~\cite{Amaricci2010,HuangLi2014,Terletska2017,Vandelli2020,Terletska2021},
what the inter-site interaction will bring to the $t^\prime$-Hubbard model 
is rarely explored at present. Here, focusing on a two-leg ladder and employing
the density-matrix renormalization group~(DMRG)~\cite{White1992,White1993,Schollwok2005,Schollwok2011}
method, we investigate the effects of repulsive interaction on the 
superconductivity in $t^\prime$-Hubbard model up to the fourth neighbor sites.
Motivated by recent experiment~\cite{ZhuoyuCHen2021}, we also explore the influences of an effective NN 
attractive interaction on $t^\prime$-Hubbard model. In the ladder systems, the numerical
errors can be well controlled, so that high
accurate numerical results are accessible.
Our numerical results indicate
that the superconducting correlation is slightly weaken but still
robust under the repulsive NN and NNN interactions. The charge density
distribution, density correlation, and spin correlations are nearly
undisturbed as the long-range repulsive interactions are considered.
In sharp contrast to the repulsive interactions, the attractive NN
interaction significantly
strengthens the superconducting correlation when the attraction is
comparable with the NN hopping $t$. The spin and density correlations show 
a nonmonotonic dependence on the attractive interaction. Before 
the attractive interaction drives the system into electron-hole phase separation~(PS), 
the coexistence of the algebraic superconducting and CDW correlations
in the ground state is consistent with the Luther-Emery~(LE) 
liquid~\cite{LutherA1974}. In addition, analyzing the different 
components of pairing correlation, we find that the pairing symmetry
tends to $d$-wave. At last, we study the impacts of on-site Coulomb 
interaction. Our numerical results show strong on-site interaction 
weakens the strength of superconducting correlation in the presence 
of both attractive and repulsive NN interactions.
These results will be shown in detail in the later sections.

The content of the paper is organized as follows: Sec. \uppercase\expandafter{\romannumeral2}
involves a brief introduction of the model and some details of DMRG
simulations. In Sec. \uppercase\expandafter{\romannumeral3},
we investigate the effects of repulsive and attractive NN interactions on the 
$t^\prime$-Hubbard model. In Sec. \uppercase\expandafter{\romannumeral4}, we probe
the long-range repulsive interactions up to the fourth-neighbors. In Sec. \uppercase\expandafter{\romannumeral5}, with  the present of 
NN repulsive and attractive interaction, we explore how the ground states affected 
by different on-site interaction $U$. This paper is closed by a summary in Sec. \uppercase\expandafter{\romannumeral6}.

\section{Model and method}

The extended $t^\prime$-Hubbard model on a two-leg ladder is written as 
\begin{equation}
    \begin{split}\mathcal{H}= & -\sum_{\langle ij\rangle\sigma}t\left(c_{i\sigma}^{\dagger}c_{j\sigma}+h.c.\right) +
        \sum_{\langle\langle ij \rangle\rangle\sigma}t^\prime\left(c_{i\sigma}^\dagger c_{j\sigma} + h.c.\right)\\
    &+U\sum_{i}n_{i\uparrow}n_{i\downarrow} +\sum_{i\neq j}V_{ij}\left(n_{i\uparrow}+n_{i\downarrow}\right)\left(n_{j\uparrow}+n_{j\downarrow}\right),\label{Ham}
\end{split}
\end{equation}

\begin{figure}[b]
    \includegraphics[width=0.45\textwidth]{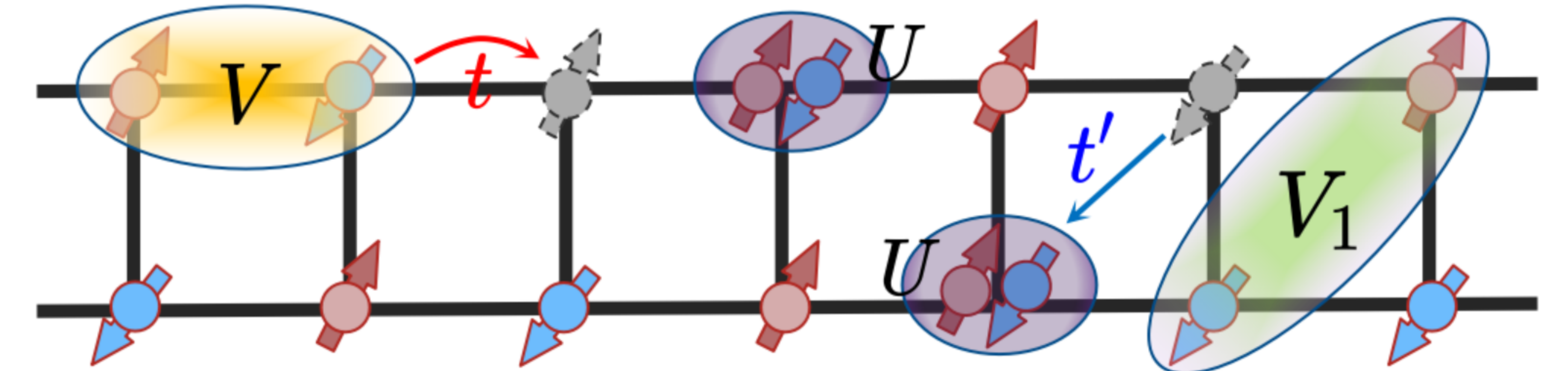} \caption{A sketch diagram of the extended $t^{\prime}$-Hubbard model on a
two-leg ladder. $t$ and $t'$ represent the nearest and the next
nearest neighbor hopping amplitudes, respectively. $U$ is on-site
Coulomb repulsion. $V$ and $V_{1}$ are the NN and the NNN inter-site
interactions. Inter-site interactions beyond NNN are not shown here.}
\label{Fig1} 
\end{figure}

where the first and second term in Eq. (\ref{Ham}) represent electrons hopping
between NN and NNN lattice sites, respectively, the coefficients $t$ and 
$t^\prime$ are their hopping amplitude. $c_{i\sigma}^{\dagger}\left(c_{i\sigma}\right)$
creates~(annihilates) an electron at site $i$ with spin $\sigma$. The third term
is on-site Coulomb repulsion for two electrons with different spins, in which $
n_{i\sigma}$ is the number of electrons with spin $\sigma$. The last term describes the 
interaction among electrons at different sites. Generally, $V$ are positive due 
to the repulsive long-range Coulomb interaction between electrons. However, the
recent experiment implies that the effective interaction between NN sites
could be attractive~\cite{ZhuoyuCHen2021}. Therefore, to investigate how
the inter-site interactions affect the ground state properties, we consider 
both repulsive and attractive inter-site interactions. For repulsive interaction,
the inter-site interactions are considered up to the fourth neighbor, and in the 
attractive case, we only consider NN interaction. We assume the values of $V_{ij}$
only depend on the distance between site $i$ and $j$. To be visually intuitive,
we illustrate the model Eq. (\ref{Ham}) in Fig. \ref{Fig1}.

In this work, we simulate the model Eq. (\ref{Ham}) by utilizing the
DMRG method, which has been shown as the most powerful method to study
one or quasi-one dimensional systems. In numerical calculations, we set 
the NN hopping amplitude $t=1$ as the energy unit, the second neighbor hopping
amplitude $t^{\prime}=-0.25$, and the on-site Coulomb interaction $U=8$ unless 
explicitly noted. These values are frequently used in related numerical simulations
for high-$T_{c}$ cuprates. The numbers of electrons with spin-up and spin-down in
$\left(\ref{Ham}\right)$ are conserved respectively, in the calculations, the two 
$U\left(1\right)$ symmetries are implemented to lower the numerical costs. We 
retain up to $8000$ states and the largest truncation error on
the order of $10^{-7}$, at least $20$ sweeps are implemented to make sure the 
calculations are well converged. The convergence of our DMRG results also checked
by the ground state energy, expectation values of observations, and the von
Neumann entropy. Our DMRG code is based on the ITensor library~\cite{ITensor}.

We adopt open boundary conditions in all the calculations. The system has
$N=2\times L$ sites and the largest system size we simulated reaches $L=96$.
At half filling, strong on-site interaction freezes the charge degree of
freedom and the ground state is Mott insulator. Hole doping makes the magnetic
order in the insulator unstable and drives the system into unconventional phases,
probably including superconductivity. The filling factor is defined
as $N_{e}/N$, where $N_{e}$ is the electron number. The concentration 
of hole is $\delta=1-N_{e}/N$. Here, we focus on the
case where the hole concentration around $\delta=12.5\%$.

\section{NN interaction}

Though some crucial characters of high-$T_c$ cuprates are 
captured by the simplest Hubbard model and $t$-$J$ model, more and more 
numerical evidences indicate the models are oversimplified and insufficient 
to explain some experimental results of cuprates. 
Some neglected subleading terms, such as
electron hopping beyond the nearest neighbor as well as inter-site interactions,
may be responsible for some puzzling novel phenomena. For example,
the NNN hopping $t^{\prime}$ induces quasi long-range superconductivity 
in the ground state~\cite{HongchenJiang2019}. In actual strongly correlated 
materials, the inter-site interactions are hard to be totally screened, and 
these interactions may also affect the properties of the system. As the NN
interaction is the most remarkable one except the on-site interaction, in this 
section, we investigate the effects of NN interaction on the $t^{\prime}$-Hubbard
model. The interactions are intuitively repulsive as they result from Coulomb 
force between electrons. However, recent experiment on a one-dimensional cuprates 
chain implies electrons in NN sites may feel an effective attraction. So, we consider 
both repulsive and attractive NN interactions.

\subsection{numerical convergency}

\begin{figure}[!htbp]
\includegraphics{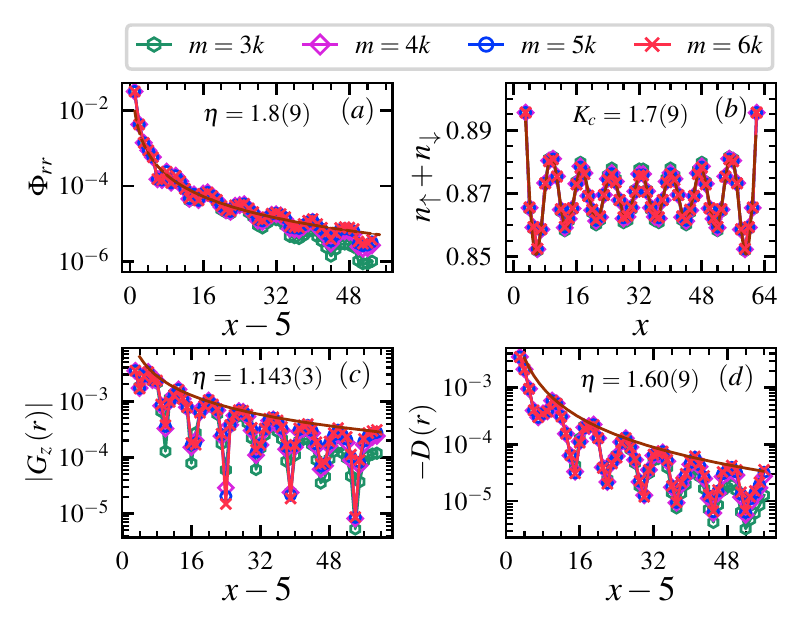} \caption{Convergency of DMRG results for $V=0.4$ as a representation. Panel
$(a)$ shows the pair-pair correlation between rungs. The brown line
in $(a)$ is a fitting via a power decay function, the solid part
represents the data used to obtain the power function, and the dashed
parts are extrapolated data from the power fitting. Panel (b) displays
the charge density profile, the brown wavy line is the fitting via the
function $n\left(x\right)=n_{0}+A{\rm cos}\left(Qx+\phi\right)$,
where $A=A_{0}\left[x^{-K_{c}/2}+\left(L_{x}+1-x\right)^{-K_{c}/2}\right]$
and Q are the amplitude and wave vector of CDW, respectively. (c)
and (d) is the spin-spin and density-density correlation, respectively,
the brown line in the two subfigures are power fittings of points on
the top of swellings. All these results indicate that 4000 retained
state is large enough to make our DMRG simulations well converged.
The common legend are shared by these subfigures.}
\label{Fig2} 
\end{figure}

Before discussing the results of NN inter-site interactions, we briefly
display the convergence of DMRG calculations and some details
in fitting the DMRG results. For simplicity, we take the results of
NN interaction $V=0.4$ as a representative sample as shown in Fig.~\ref{Fig2},
the convergence of the calculations under other parameters are similar.
To diagnosis the possibility of superconductivity, the singlet pairing
correlation is defined as

\begin{equation}
\Phi\left(x\right)=\left\langle \Delta^{\dagger}\left(x_{0}\right)\Delta\left(x_{0}+x\right)\right\rangle .
\end{equation}
Here, the variable of the pairing field denotes the position of a
bond on leg or rung. For example, when we consider the pairing correlation
between rungs, the spin singlet pair-field creation operator $\Delta^{\dagger}\left(x\right)$
is given by $\Delta^{\dagger}\left(x\right)=\frac{1}{\sqrt{2}}\left[c_{\left(x,0\right),\uparrow}^{\dagger}c_{\left(x,1\right),\downarrow}^{\dagger}-c_{\left(x,0\right),\downarrow}^{\dagger}c_{\left(x,1\right),\uparrow}^{\dagger}\right],$
where the site index is labeled by the rung index $x$ and leg index
$y=0,1$, respectively. Similarly, the charge density profile is $n\left(x\right)=1/2\left(n_{x,0}+n_{x,1}\right)$,
$S_{z}$ component of spin correlation $G_{z}\left(x\right)=\left\langle S_{x,i}^{z}S_{x_{0},i}^{z}\right\rangle $,
and charge density correlation $D\left(r\right)=\left\langle n_{x,i}n_{x_{0},i}\right\rangle -\left\langle n_{x,i}\right\rangle \left\langle n_{x_{0},i}\right\rangle $.
The number of retained states $m$ ranges from $3000$ to $6000$
with an increment of $1000$. When the number is greater than or equal
to $4000$, all the results have a good convergence such that their
differences are tiny as shown in Fig.~(\ref{Fig2}). Therefore, $4000$ retained
states are sufficient to obtain accurate numerical results, and there is
no need to extrapolate the DMRG results to the $m=\infty$ limit. In the later 
DMRG calculations, we keep at least $4000$ states to ensure the numerical
accuracy. The brown lines in Fig.~(\ref{Fig2}) are the fittings of corresponding data.
Power decay functions can fit the correlation functions well, and the charge
density profile can be fitted via a trigonometric function multiplied
by a spacing-dependent amplitude given in Fig.~\ref{Fig2}. The spin
and charge correlations behave in a spatial oscillation with considerable
amplitudes, so we use the points on the top of each upward bulge to
obtain the power function. In the following discussion, we will not
involve the details of fittings any more and directly use the power
exponent extracted from the power function.

\subsection{repulsive interaction}

\begin{figure}[!htbp]
\includegraphics{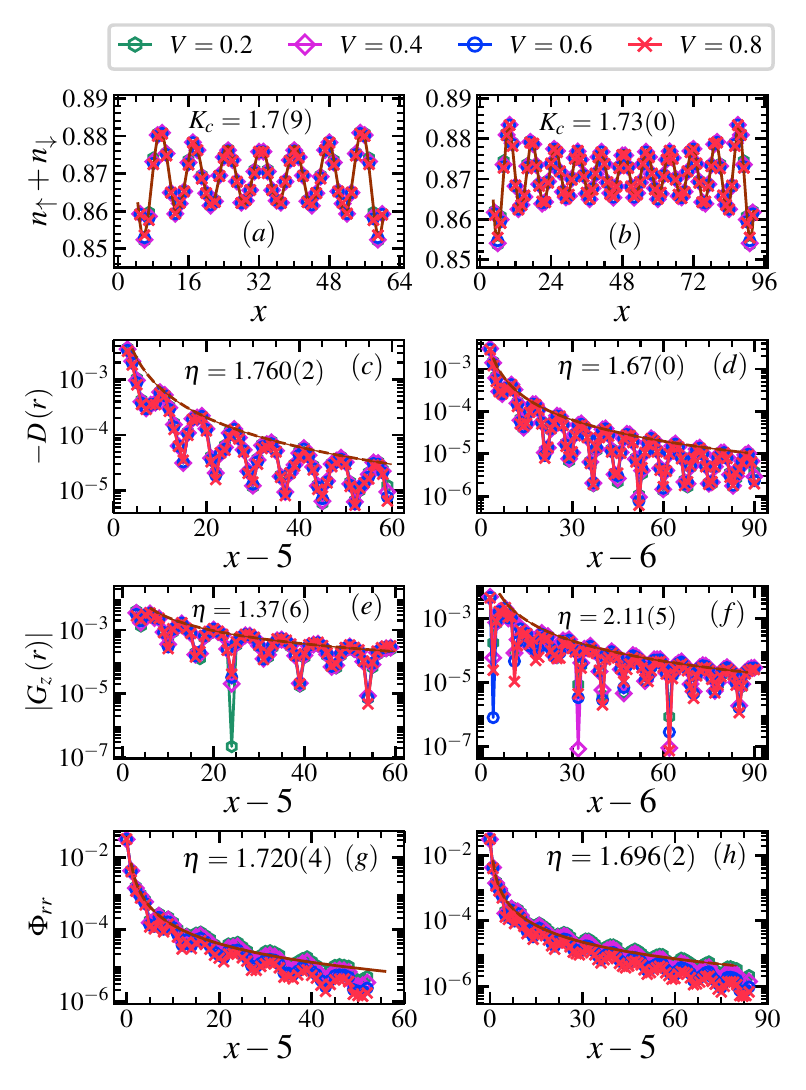} \caption{ The charge density profile, density correlation, spin correlation,
and superconducting correlation under different strengths of repulsive
NN interaction, $V$ ranges from $0.2$ to $0.8$ with an increment
of $0.2$. The left column is the results of $L=64$, the results
of $L=96$ are shown in the right column, the DMRG results on these
two system sizes are well consistent.  The brown lines in these figures 
are power fittings of the results of $V=0.2$.  In $\left(a\right)$ and 
$\left(b\right)$, the points close to boundaries are dropped, but all the 
valleys and peaks are retained, form which we can easily obtain the wavelength
of CDW. The $y$-axes in the subfigures in each row share the same
labels. All these subfigures shares the same legend. }
\label{Fig3} 
\end{figure}

Next, we add the NN repulsive interaction. Fig.~\ref{Fig3} shows 
the effects of repulsive $V$ term on the charge density profile, density 
correlation, spin correlation, and the superconducting correlation between 
rungs. The strength of $V$ exhibited here ranges from $0.2$ to $0.8$ with
a step of $0.2$, the left column and right column in Fig.~\ref{Fig3}
correspond to the results of two different system sizes $L=64$ and
$L=96$, respectively. The quantities shown here share the same behaviours
for the two sizes, so the finite size effects are minor and should
not qualitatively alter our conclusions. Firstly, as shown in Fig.~\ref{Fig3}
$\left(a\right)$, $\left(c\right)$, and $\left(e\right)$, the charge
density profile, density correlation, and spin correlation are insensitive
to $V$. Under various $V$, each of the behaviours can be approximately
fitted by a single function. The CDW is robust in the presence of
NN Coulomb interaction, only the charge densities close to the boundaries
are slightly affected as we adopt the open boundary conditions. Clearly,
from Fig.~\ref{Fig3} $\left(a\right)$ and $\left(b\right)$ the
wavelength of charge density is $\lambda_{c}=8$, and there is one
hole in each stripe. Analyzing the spin correlation function, we find
that there are domain walls in the antiferromagnetic background where
holes are enriched. The period of spin order $\lambda_{s}$ is about
$16$, twice of the charge density wave. Secondly, the last row subfigures
$\left(g\right)$ and $\left(h\right)$ in Fig~\ref{Fig3} indicate
that the superconducting correlation is susceptible to the repulsive
NN interaction. For the both system sizes, a larger power exponent of
the  function is required to fit the superconductivity
correlation function when the strength of repulsive NN interaction
is increased. Thus, the superconductivity is weakened by repulsive
NN Coulomb interaction. At last, the numerical results show that the
superconductivity is stable against the repulsive NN interaction,
though the NN interaction weakens the superconducting correlation,
as shown in Fig.~\ref{Fig3}, instead of damaging the superconductivity.
For larger $V$, even if $V=1.6$ (not shown), the superconductivity
correlation still obeys a power decay function.

For the $t^{\prime}$-Hubbard model with only on-site interaction,
the Luttinger parameter $K_{c}$ of CDW, extracted from the CDW function
given in Fig.~\ref{Fig2}, is comparable to the Luttinger parameter
of superconductivity or the decay exponent of superconducting correlation.
The superconducting correlation is weakened by NN repulsive interaction. 
Though superconductivity and CDW coexist in the ground state, with the 
increase of repulsive NN interaction, the CDW dominates in the ground state.

\subsection{attractive interaction}

The numerical results in above indicate that the ground state 
is not inclined to superconductivity at the presence of repulsive
inter-site interactions. Recently,  some experimental spectra 
characteristics of a one-dimensional cuprates chain can be well understood
via the Hubbard model with attractive NN interaction. 
Furthermore, numerical simulation have identified that an effective attractive 
interaction is induced in the Holstein-Hubbard model with long-range 
electron-phonon coupling. The experimental and  numerical work renewed 
people’s interest in Hubbard model with attractive NN 
interaction~\cite{JiangMi2022,QudaiWei2022,PengCheng2022, TangTa2022, WangHaoxin2022}.
In the one-dimensional chain, the ground state exhibits $p$-wave
superconductivity~\cite{QudaiWei2022}. And the correlation of $d$-wave
superconductivity is enhanced by attractive interaction on a four-leg
cylinder~\cite{PengCheng2022}. Here, we consider the Hubbard model
with attractive NN interaction on the two-leg ladder. Through extensive
numerical simulations, we give more insights on how the spin correlation,
density correlation, and the superconducting correlation are affected
by the attractive interaction. Cooperating with recent numerical simulations,
we can gain a deeper understanding of the Hubbard model with attractive
interaction.

\begin{figure}[t]
\includegraphics{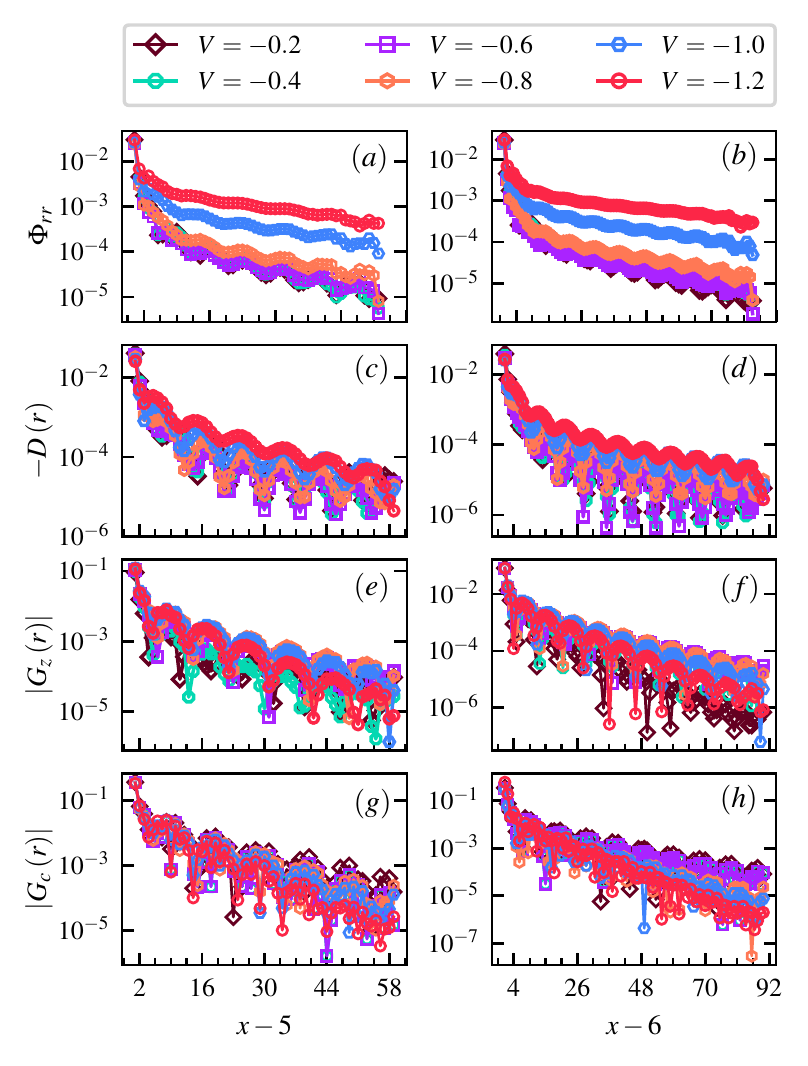} \caption{The left column and the right column are the results of system sizes
$L_{x}=64$ and $L_{x}=96$ under different attractive NN interaction,
respectively. The numerical results of the two system sizes are well
consistent. $\left(a\right)$ and $\left(b\right)$ are the superconductivity
correlation, it is enhanced with the increase of $V$. The enhancement of 
superconductivity becomes remarkable when the strength of $V$ is larger than $0.6$.
The density correlation function, shown in panels $\left(c\right)$ and $\left(d\right)$,
always decays algebraically. In pasnnels $\left(e\right)$ and $\left(f\right)$, 
the spin correlation changes from a algebraic decay into an exponential decay 
with the increase of $V$, indicating the open of spin gap. $\left(g\right)$ and
$\left(h\right)$ show the single particle Green function,
it decays algebraically under small $V$, but exhibits an exponential decay
as $V$ becomes strong, which implies the open of charge gap when strong attractive $V$ is
introduced. The subfigures $\left(a\right)$-$\left(h\right)$ use
the same legend.}
\label{Fig4} 
\end{figure}

\begin{figure}[t]
\includegraphics{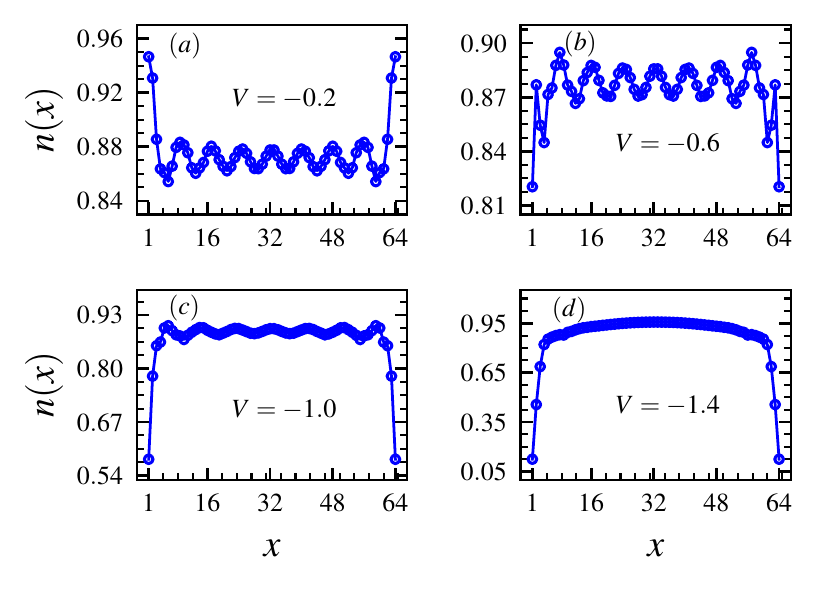} \caption{Typical charge density distributions under different $V$ with the
system size $L_{x}=64$. Since open boundary condition are implemented,
electrons prefer to accumulate at the boundaries for small $V$, while
holes on the boundaries are favorable when $V$ is large. If $V$
is not strong enough, as show in $\left(a\right)$, $\left(b\right)$,
and $\left(c\right)$, the modulation of change density is maintained.
In the panel $\left(d\right)$, The modulation is destroyed by strong
NN attractive interaction, electrons are gather in the bulk and the
holes are at the two edges, indicating a phase separation.}
\label{Fig5} 
\end{figure}

Following the case of NN repulsive interaction, we also take two system sizes
$L=64$ and $L=96$ to show the numerical results. The main results are 
summarized in Fig.~\ref{Fig4}. The numerical results on the two systems 
sizes are well consistent. The superconducting correlations for the two 
systems are shown in Fig.~\ref{Fig4} $\left(a\right)$ and $\left(b\right)$.
When the strength of $V<0.6$, the attractive interaction has no obvious 
effects on the superconducting correlation. With increasing of $V$, 
the superconducting correlation shows a slower and slower decay, indicating
a significant enhancement by strong attractive $V$. The spin correlation is 
enhanced by weak $V$ while weakened by strong $V$. Meanwhile, the decay of 
the weakened spin correlation changes from a power law to an exponential law, 
which indicates the spin order is quenched by the strong attractive interaction.
The charge density correlation is slightly weakened by small $V$ but it is
strengthened by strong $V$, which is a contrary to spin correlation.
The different dependencies of spin correlation and superconducting correlation on $V$ 
indicates that a competition exists between them. For large $V$, the valleys in density correlation
become shallow, as shown in Fig.~\ref{Fig4}, and the charge density
distribution cannot be fitted by the function for CDW used in Fig.~\ref{Fig2}
at large $V$. The exponential decay of spin correlation and
single particle Green's function show that both the spin and charge
excitations are gapped. In addition, the density correlation obeys
a power decay behaviour. These indicate large $V$ drive the system
into an LE liquid phase. After the $V$ drive the system into the
LE liquid, the superconductivity is significantly enhanced. For very
strong NN attractive interaction, the system is driven into electron-hole
SP, which is characterized by regions with rich holes and
regions with rich electrons. Due to the attractive NN interaction
and OBC in our DMRG simulation, the holes prefer to accumulate in
the two edges for strong attraction, while the electrons tend to gather
in the inside. The evolution of charge density distribution 
under different attractive $V$ is shown in Fig.\ref{Fig5}.

\subsection{pairing symmetry}

\begin{figure}[t]
\includegraphics{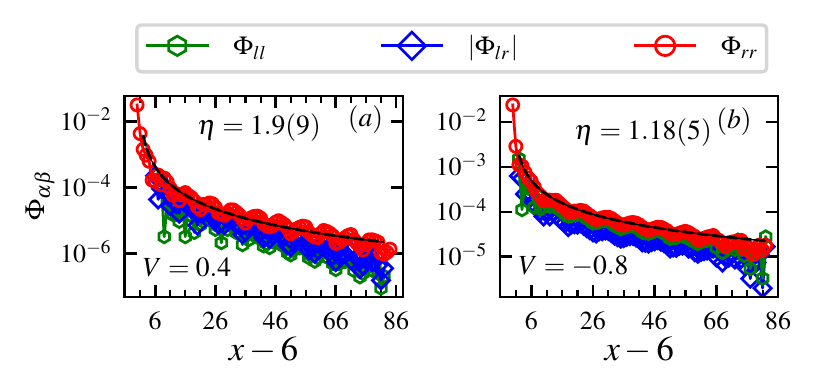} \caption{Three different superconducting correlations, the singlet pairing
correlation between a reference rung bond and bond on rungs $\Phi_{rr}$,
and bonds on one leg $\Phi_{rl}$, as well as two bonds on the same
leg $\Phi_{ll}$, for $V=0.4$ and $V=-0.8$ with $L=96$ are shown. Since the
constraint of ladder geometry, we only show the long range part of
$\Phi_{rl}$ and $\Phi_{ll}$.
The long-range behaviors of the correlations in the two cases tell
us $d$-wave superconductivity are favored in the ground state. These
two subfigures share the same legend.}
\label{Fig6} 
\end{figure}

At the end of this section, we briefly discuss the pairing symmetry
on the two-leg ladder. Though the two-leg ladder does not have the 
same spatial symmetry in the two directions along leg and rung, through
pairing correlations we may still gain some sights about the pairing 
in the two-dimensional case. Here, three different superconductivity 
correlations, $\Phi_{rr}\left(x-x_{0}\right)$, $\Phi_{rl}\left(x-x_{0}\right)$ 
and $\Phi_{ll}\left(x-x_{0}\right)$, are used to diagnosis the pairing symmetry,
where the index $r$ and $l$ represent the bond in the rung and leg respectively. 
The first index represents the reference bond at position $x_{0}$, and
the second index represents the other bond at $x$. The superconductivity
correlation for two typical $V$ are shown in Fig.\ref{Fig6}, they
satisfy the characters of $d$-wave symmetry. These results mean that
for the $t^{\prime}$-Hubbard model, with the presence of repulsive
or attractive NN interaction, the superconductivity tends to be $d$-wave.
When the strength of NN attractive interaction is comparable to NN
hopping $t$, the $d$-wave superconductivity is significantly enhanced.
The dependency of superconductivity on the strength of NN attractive
is consistent with recent DMRG simulation on a four-leg square cylinder~\cite{PengCheng2022},
which confirm that a $d$-wave superconductivity may be stabilized
by NN interaction in two-dimensional square lattice.

\section{long-range inter-site interaction}

\begin{figure}[!htbp]
\includegraphics{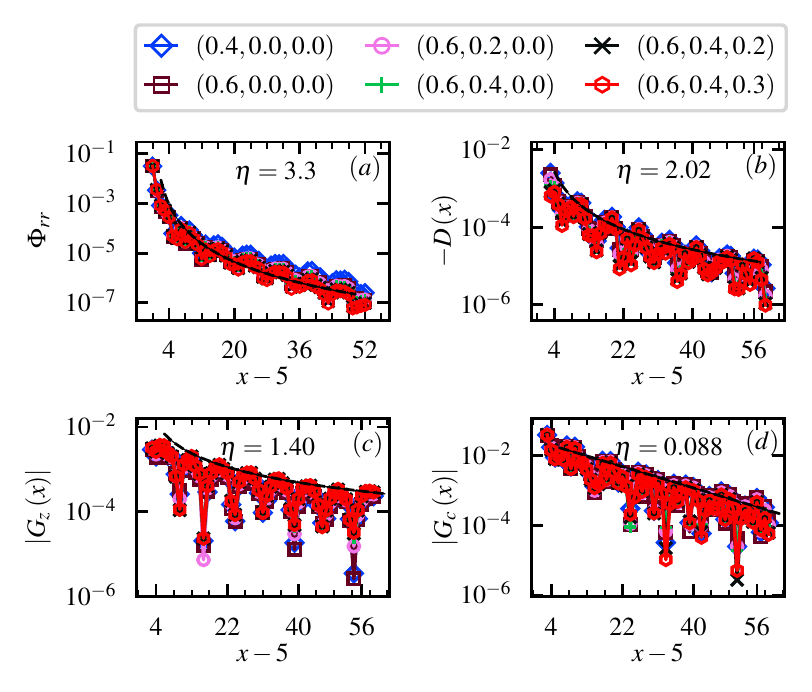} \caption{Correlation functions when repulsive inter-site interactions beyond
the NN neighbor are taken into account. The legend represent the values of inter-site
interaction in the form of $(V_1, V_2, V_3)$, which are shared by these subfigures.
Panel $\left(a\right)$ is the superconducting correlation between rungs, the 
repulsive NNN interaction $V_{1}$ will slightly weaken the correlation. For the
third neighbor and the fourth neighbor interactions, the superconductivity
correlations are nearly the same. Meanwhile, the superconductivity
correlation is hard to be fitted by an exponential decay function,
the power decay functions gives a better fitting. The spin correlation
function, density-density correlation, and single particle Green function,
shown in $\left(b\right)$, $\left(c\right)$, and $\left(d\right)$
respectively, are insensitive to these interactions, both the spin
correlation and density correlation obey power law decay. But the
single particle Green function decays exponentially.}
\label{Fig7} 
\end{figure}

Now we consider the inter-site Coulomb interaction beyond the NN sites,
ranging to the fourth neighbor sites. In following calculations, we
set NN interaction $V=0.8$. The results of $L_{x}=64$ are shown
in Fig.~\ref{Fig7}. The superconducting correlation is shown in
panel $\left(a\right)$ of Fig.~\ref{Fig4}. With the increase of
the NNN interaction form $V_{1}=0.4$ to $V_{1}=0.6$ the superconducting
correlation is suppressed slightly. With $V_{1}=0.6$ and the long-range 
third and fourth neighbor interactions involved, the superconducting correlation
nearly remains unaffected as shown in the legend of Fig.~\ref{Fig7} in the form of $(V_1, V_2, V_3)$ 
for the specific values of the interactions strength. Just like the case of NN interaction,
the power decay function gives a better fitting of the superconducting correlation
than the exponential function, indicating the robustness of the superconductivity
correlation to the complex interactions. In the whole process, the
charge density profile (not shown), density correlation, spin correlation,
as well as the single-particle Green function are insensitive to these
repulsive interactions. If we take a closer look, the strengths of density 
correlation and single particle's Green function are subtly suppressed by these long
range interactions, while these long range inter-site interactions marginally enhance
the strength of spin correlation.  Except the single-particle Green function,
which decays exponentially, both spin and density correlations decay
algebraically.

Based on these simulations, we can conclude that the ground state
properties of $t^{\prime}$-Hubbard model are quite robust to these
long-range repulsive inter-site interactions. Though the inter-site repulsion
between electrons tends to weaken the superconductivity, the algebraic
superconducting correlation is not destroyed by the inter-site interactions,
even if the strengths of these interaction are comparable or much stronger
than the strength of $t$. Since these repulsive inter-site
interactions always tend to weaken the superconductivity, reducing
these repulsive interactions should be beneficial for the superconductivity.

\section{variation in on-site Coulomb interaction}

\begin{figure}[!htbp]
\includegraphics{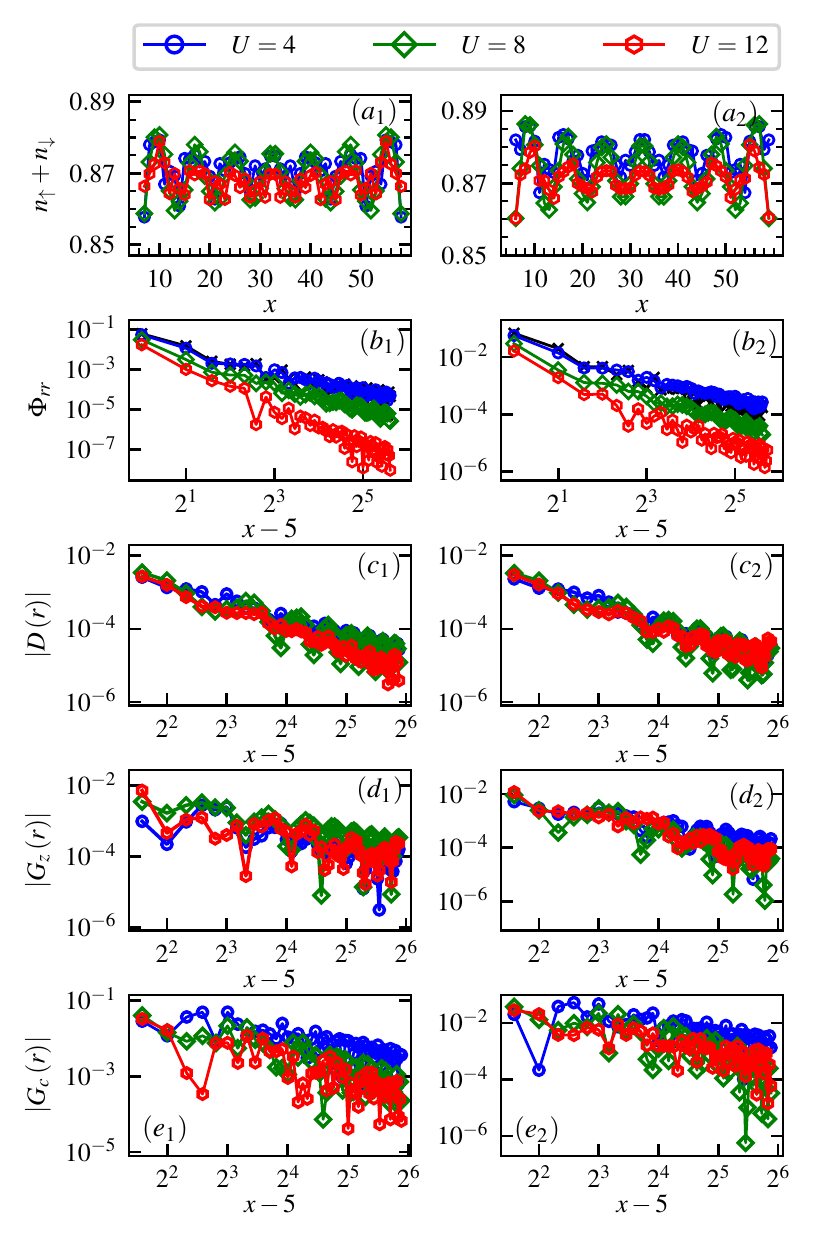}
\caption{When the system size $L=64$, the left column and the right
column are the results under different Hubbard $U$ when the NN interaction
$V=0.4$ and $V=-0.4$, respectively. Subfigures $\left(a_1\right)$ and
$\left(a_2\right)$ are the charge density distributions, the CDW order is stabilized
under medium-strength of $U$, where $U=6, 8, 10$. $\left(b_1\right)$ and 
$\left(b_2\right)$ show the superconducting correlation, large $U$ 
tend to weaken the strength of the correlation. Under same $U$, the strength
superconductivity correlation of repulsive $V$ is weaker than the one of
attractive $V$, the data in black is the results of $U=2$. The density 
correlations are given in $\left(c_1\right)$ and $\left(c_2\right)$, they decay
algebraically and slightly weakened overall with the increment of $U$, 
and it is easier weakened under repulsive $V$. The spin correlations are shown 
in panels $\left(d_1\right)$ and $\left(d_2\right)$, they are  enhanced
at first and suppressed when $U$ is large, the spin correlations in the two cases
satisfies a power decay. In the last row, $\left(e_1\right)$ and $\left(e_2\right)$,
the single particle Green's  functions are suppressed with the increase of $U$, 
it is easier to be weakened  when increasing $U$ for repulsive $V$ than for 
attractive $V$, the legend are shared by all the subfigs, for clarity, the data 
of $U=2, 6, 10$  are not shown.}
\label{Fig8} 
\end{figure}

At last, we investigate the effects of on-site Coulomb interaction.
On the one hand, when electrons in materials are less localized, the 
band width can be larger so that a smaller relative $U$ should be considered. On the other hand, people are also
interested in the physics of large $U$ limit of Hubbard model under doping.
For simplicity, considering both repulsive and attractive NN interaction 
with $|V|=0.4$, we vary the strength of on-site Coulomb interaction from $2$
to $14$ with an increment of 2, all the other parameters in the extended 
Hubbard model are fixed. For clarity, we mainly show the results of $U=4, 8, 12$
in Fig.~\ref{Fig8}.

Despite some minor differences in the two cases, the charge density profile,
and the correlations shown in Fig.~\ref{Fig8} share many common behaviors 
on the dependency of the Hubbard $U$, which are expected to the properties
of $t^\prime$-Hubbard model itself. Firstly, the superconducting correlation 
is enhanced as $U$ is reduced and tends to be saturated. For 
example, the strength of superconducting correlation of $U=4$ is almost 
the same as the one of $U=2$. Secondly, the wavy charge density distribution 
is supported when $U$ is around $8$, while under both stronger and smaller $U$, it 
becomes unstable,the amplitude of inhomogeneous distribution reduced and the wavy feature
disappeared. The instability under very large $U$ agrees with a previous 
result of Hubbard model on a four-leg cylinder~\cite{JiangYifan2020}. The 
CDW collapses when $U=12$ for repulsive $V$, but it still 
exists for attractive $V$ and cracks until we further increase $U$. These 
indicate the CDW under attractive $V$ is more stable against the interference
of large $U$. The CDW also destabilizes when a smaller $U$ is considered. Though the 
charge density undulates, it starts getting close to a uniform distribution. Our 
results imply that uniform superconductivity is likely to form under small $U$.
Here, the numerical results give the tendency that the superconductivity and 
stripe is not tightly related in the $t^\prime$-Hubbard model on a two-leg ladder.
The superconducting correlation is established at small $U$, though increasing 
$U$ weakens its strength, the power decay behavior is robust in the whole range of
$U$ we considered here. CDW is induced until $U$ reaches certain strength
and it is damaged at large $U$.

Under the two situations, the density-density correlations, spin correlations 
as well as the single particle Green's functions obey a power law decay 
behavior, implying the spin excitation and charge excitation are gapless.
The density correlation is slightly suppressed with the increase of $U$.
Compared with attractive $V$, the density-density correlation are easier 
weakened under repulsive $V$. The spin correlation is enhanced at first and then 
is suppressed when $U$ increases,i and it has a wilder window of
enhancement for repulsive $V$. The single particle Green's function is overall
weakened as we increase $U$, it is more sensitive under repulsive $V$.

\section{conclusions}

Using the DMRG method, we have systematically investigated how the
complex interactions affect the ground state of the $t^{\prime}$-Hubbard
model on a two-leg ladder. Without inter-site
interactions, the superconducting order and the CDW order are comparable.  The NN repulsive interaction weakens the superconductivity
order and leads to the domination of CDW. When the long range Coulomb interaction
is well screened it is more likely
to induce superconductivity. Our numerical results show
that the ground state properties are stable against on the longer-range repulsive
interactions. The exact numerical results clearly indicate
repulsive interactions between different sites are adverse to the
superconductivity. However, the superconductivity
in $t^{\prime}$-Hubbard is not damaged by physical reasonable repulsive
inter-site interactions on a two-leg ladder. 

Motivated by recent experimental results as well as the numerical
simulations, we also investigated the attractive NN interaction on
a two-leg ladder. The numerical results show that both the spin order
and charge order are sensitive to the attractive interaction, though
the superconductivity is less affected when $V$ is small. But for
relatively large $V$ the superconductivity order is significantly
enhanced. For very large $V$, the system evolves into phase separation.
Before entering the phase separation, we find an LE liquid phase.
Our numerical results show strong evidence that the superconductivity
is strengthened by strong attractive NN interaction, and it is considerably
enhanced when the strength of $V$ is comparable to the NN hopping
amplitude $t$, agreeing with a recent DMRG study on a four-leg cylinder.

When the superconductivity is quasi-long ranged, by analyzing the
different singlet pairing correlations, we find the superconductivity
tends to be $d$-wave. Both in repulsive and attractive interactions,
the superconductivity is unfavorable under strong Hubbard interaction
$U$. In addition, the charge density wave is supported when $U$
is around $8$, both weaker and stronger $U$ make the charge density
unstable.

We have extensively explore the effects of complex inter-site interaction
on the superconductivity of $t^{\prime}$-Hubbard model on a two-leg
ladder. As there exist superconducting cuprates ladder materials, our
numerical results can be helpful in understanding the novel physics
in these materials. Ladder serves as a bridge between one and two
dimensional systems, our results can also shield some lights on revealing
the high-$T_{c}$ superconductivity in cuprates.

\textbf{Acknowledgments:} 
{This work is supported by the National Key Research and Development Program
of China (Grant No. 2022YFA1402704) and by the National Natural Sience Foundation 
(Grants No.12274187, No.12047501, No.12274187, No.12247101, No. 11834005)}.

%

\end{document}